      [1999/12/01 v1.4c Il Nuovo Cimento]
\documentclass{cimento}

\usepackage{graphicx}  
\title{Cosmological model: \\ from initial conditions to structure formation}
\author{V.N.~Lukash\from{ins:x}\thanks{lukash@asc.rssi.ru}}
\instlist{\inst{ins:x} Astro Space Centre of P.N. Lebedev Physics
Institute - Profsoyuznaya 84/32, Moscow 117997 Russia}
\PACSes{\PACSit{95.35.+d, 98.65.Dx, 98.80.-k}{}
\PACSit{---.---}{\ldots}}
\begin{document}

\maketitle

\begin{abstract}
Observational cosmology is on the verge of new discoveries that
will change the essence of our world-view. The matter concerns
origin of initial conditions and physics of dark matter.
\end{abstract}

\section{Identification problem}

A century of cosmology has led us to a new understanding of the
Universe. Today we know the model at large scales. {\it Per aspera
ad astra}. After many years of hypotheses and markets of models we
now have the standard cosmological model, yet separated from what
we have at small scales -- the standard model of elementary
particles. Both models progressively converge and interact with
each other leading us to a joint physical model of the World we
are a part of.

The progress in cosmology is ensured by observations. This creates
identification problem. It is a specific feature of astronomy.
Astronomers \textit{see} structures unknown to physicists. They
cannot touch or test them, they can learn only general properties
of observed matters assuming some theoretical extrapolations
(General Relativity, atomic physics, etc.). On the contrary,
physicists need experiment to judge things. To understand what
astronomers see, physicists are looking in labs for what is
unknown to them, since there is not enough information about the
target. In this way the problem of identification arises.

What do astronomers see?

They observe structures made of invisible matter, the {\it dark
matter}. DM does not interact with light, generally - with
luminous matter, or baryons. How is DM observed then? Through its
gravitational influence on visible matter. Fortunately, light is
there where DM concentrations are.

Fig. 1 shows a region of sky in the direction of one of DM {\it
halos}, the non-linear DM concentration gravitationally bound in
all three directions with total mass $\sim 10^{14}M_{\odot}$. We
see optical galaxies captured by gravitational field of this
concentration, X-ray gas residing at the bottom of the
gravitational well, and a multi-image of one of the background
galaxies that happened to be on the line of sight of the DM halo
and was distorted by its gravitational field.

\begin{figure}
\includegraphics{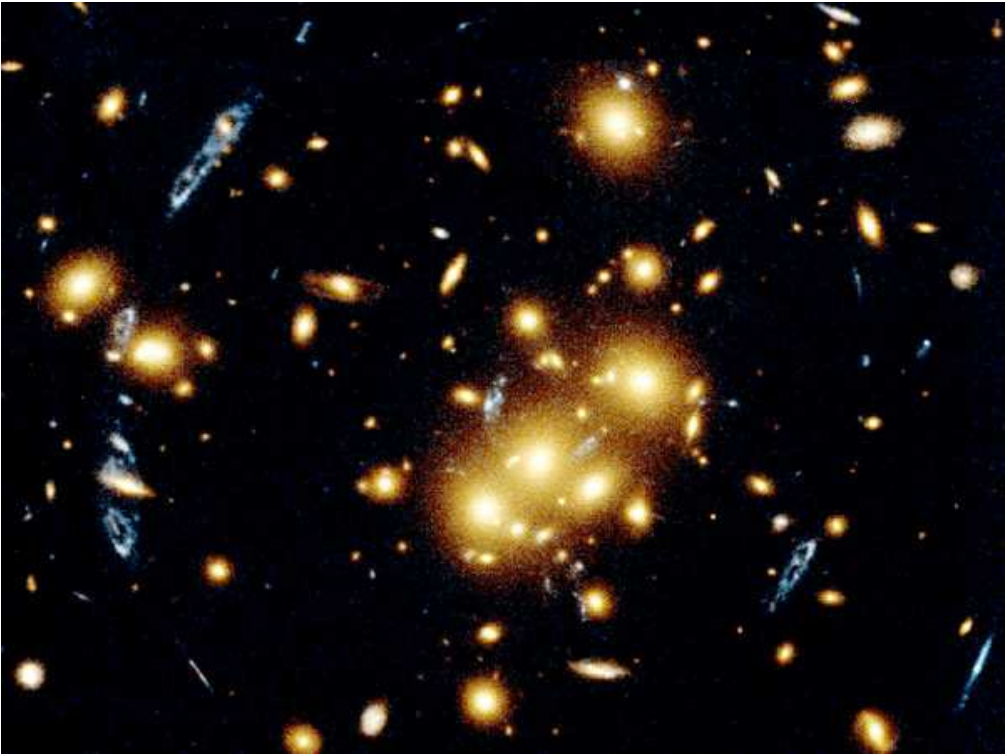}     
\caption{Hubble Space Telescope photo of a sky region in the
direction of cluster of galaxies $0024+1654$.}
\end{figure}

We study spatial distribution of DM halos analyzing galaxy
catalogs and quasar absorption lines. Besides, the DM surface mass
density can be reconstructed via its gravitational week lensing
action on numerous background galaxies. So, there is more than
enough independent probes of dark mass inside and beyond DM halos.
We can state that the mean contrast of DM density field is larger
than unity at small scale ($< 10$Mpc) still remaining less than
unity at large scale ($\ge 10$ Mpc). Accordingly, we do not find
DM halos exceeding $10^{15}M_{\odot}$.

This is the current DM density field. We are lucky to have a map
of much younger matter density field using CMB anisotropy. That
time ($z\sim 1000$) the mean density contrast was $\sim 10^{-5}$,
and no halos had formed yet. Having these two pictures of cosmic
matter distribution at different epochs of its evolution and
assuming that only gravity is responsible for such evolution, we
can obtain the DM energy-momentum tensor.

What are DM properties?

Actually, they are very simple. DM is non-relativistic {\it weakly
interacting massive particles} with cosmological density five
times higher than that of baryons. WIMPs should be cold
(non-relativistic) long before the equality epoch to be able to
form the structure that we observe today. Owing to such simple
properties, DM has straightforwardly affected the development of
the Universe gravitational potential. The DM density contrast was
increasing in time due to gravitational instability. Baryons,
after they decoupled from radiation, were captured into
gravitational wells of DM concentrations. That is why light is
there where DM is, although DM particles do not interact with
light. Thanks to this remarkable feature of gravitational
instability it is possible to study amount, state and distribution
of DM in observations ranging from radio to X-ray bands.
Contemporary physics does not know particles with DM properties.
It is necessary to go beyond the standard model. But how and in
which direction? What should we look for?

The analysis of large scale structure in the Universe has revealed
that the amount of non-relativistic DM entering structure is
small. The overall mass density of all particles which have been
involved in the process of gravitational clustering, cannot exceed
$30\%$ of the critical density. Massive neutrino contribute less
than few percent to the matter budget. At the same time the
characteristics of CMB anisotropy have evidenced the flat spatial
geometry of our Universe. It means that the rest $70\%$ of the
critical density should be in the form that takes no part in
gravitational clustering. What are the properties of such a stable
medium which is not perturbed by gravitational potential of the
structure and remains essentially unclustered?

Theory gives a clear answer to this question -- the
pressure-to-energy ratio of this medium, $w\equiv p_{\rm
DE}/\epsilon_{\rm DE}$, should satisfy the following condition:
\begin{equation}
|1+w| << 1\;.
\end{equation}
Only under this inequality the medium remains Lorentz-invariant
and invariable both in space and time. We call it {\it dark
energy}. This is all we know about DE.

It is crucial that the process of gravitational instability could
be launched in the Friedmann Universe only if the seed density
perturbations were present since the very beginning. The existence
of primordial cosmological perturbations has nothing to do with DM
or any other particles. These are the total density perturbations
that were produced by the Big Bang physics. Thus, another
important problem arises, the problem of origin of the seed
density perturbations which have developed dynamically into DM
structures.

These hot topics -- searching for unknown matter and determining
the initial conditions for structure formation -- display new
physics and are expected to be solved in near future. In this
review we dwell upon them.

\section{Geometry of late and early Universe}

The observed structure of the Universe is a product of start
conditions and evolution of matter density field. Up-to-date
observational data made it possible to determine characteristics
of the density field at different epochs of its development. It
allowed us to separate information about the initial conditions
and development conditions, thus giving rise to independent
investigations of the early and late Universe physics.

In modern cosmology the term "early Universe" stands for the final
period of the inflationary {\it Big Bang stage} with subsequent
transition to hot period of cosmological expansion. Currently we
have no model of the early Universe as we do not know BBS
parameters (there are only upper bounds, see eq.~(\ref{eq-12})).
However, we have a well-developed theory of quantum-gravitational
generation of the cosmological perturbations. Using this theory,
we can derive the spectra of primordial density perturbations and
cosmic gravitational waves as functions of cosmological
parameters, and constrain the latter if the spectra are known.

The reason why we still have no generally-accepted model of the
early Universe stems from stable predictions of BBS inflationary
paradigm, that are obtained in a wide class of model parameters.
Namely, the generated spectra are almost flat, the amplitude of
the cosmological gravitational waves is relatively small, etc.
Detection of the cosmological gravitational waves would give
crucial information about the early Universe. This discovery may
come in case the PLANCK experiment succeeds.

Our knowledge of the late Universe is quite opposite. We have
rather precise model -- we know the main matter components and
cosmological parameters, the evolution of the Universe and theory
of structure formation. But we do not understand how the matter
components have originated.

The known properties of the visible Universe allow us to describe
the geometry of both, late and early Universe, in the framework of
the perturbation theory. There is a small parameter $\sim
10^{-5}$, the amplitude of cosmological perturbations.

The main tool of geometry is metric tensor. To zeroth order the
Universe is Friedmannian and described with only one time function
-- the scale factor $a(t)$. The first order is a bit more
complicated. The metrics perturbations are the sum of three
independent modes --the scalar one $S(k)$, the vector one $V(k)$,
and the tensor one $T(k)$, each of them being described by its
spectrum, the function of the wave number $k$. The scalar mode
describes the cosmological density perturbations, the vector mode
is responsible for vortical matter motions and the tensor mode
presents the gravitational waves. If the first order fields are
Gaussian then the entire geometry of our Universe is described
with only four positively defined functions, $a(t)$, $S(k)$,
$T(k)$ and $V(k)$. Currently we know the first two functions in
some ranges of their definition.

BBS was a catastrophic process of rapid expansion accompanied by
intensive time varying gravitational field. Under this
gravitational action the small-scale cosmological perturbations of
metric and density were being parametrically born from vacuum
fluctuations. It is very general and fundamental effect of
creation of any massless degree of freedom in external coupled
non-stationary field.

Available observational data confirm the quantum-gravitational
origin of seed density perturbations responsible for structure
formation in the Universe. It is a good example of the solution of
measurability problem in quantum field theory. The basic
properties of the perturbation fields generated according to this
mechanism are the following: the Gaussian statistics (random
distribution in space), the preferred time phase ("growing" branch
of evolution), the absence of characteristic scales in a wide
range of wavelengths, a non-zero amplitude of the gravitational
waves. The latter is crucial for building-up the BBS model as
gravitational waves couple the simplest way to the background
scale factor.

The development of $S$-mode has resulted in formation of galaxies
and other astronomical objects. The CMB anisotropy and
polarization have emerged long before under the joint action of
all three perturbation modes ($S, T$ and $V$) on the photon
distribution. Joint analysis of the observational data on galaxy
distribution and the CMB anisotropy allowed us to relate $S$ and
$T+V$ modes. Making use of the fact that the sum $S+T+V\simeq
10^{-10}$ is known from the CMB anisotropy, we obtain the upper
bound for the vortical and tensor perturbation modes in the
visible Universe:
\begin{equation}
\frac{T+V}{S}<0.2 \label{eq-1}
\end{equation}
In case the latter inequality were violated the density
perturbation value would not be sufficient to form the observed
structure. The detection of $T$ and/or $V$ (e.g. cosmological
magnetic field) will become possible only with increase of
observational precision.

\section{Is DE a massive field ?}

Let us consider zero order geometry more detailed.

Table \ref{tab:table1} presents average values of the cosmological
parameters obtained from astronomical observations (with $10\%$
accuracy). With these parameters, we obtain from the Friedmann
equations the Hubble function, $H\equiv{\dot a}/a$, and its time
derivative, $\gamma\equiv-{\dot H}/H^2$:
\begin{eqnletter}
\frac{H}{H_0}=10^{61}\frac{H}{M_P}=\left(
\frac{10^{-4}}{a^4}+\frac{0.3}{a^3}+0.7\right)^{1/2}\;, \\
\gamma=\frac{d\ln(M_P/H)}{d\ln a}= \frac{3(\epsilon+p)}{2\epsilon}
=\frac{2\cdot 10^{-4}+0.4 a}{10^{-4}+0.3 a+0.7 a^4}\;,
\end{eqnletter}
where $H_0^{-1}= 14 {\rm Gyr}=10^{33}{\rm eV}^{-1}$ is the inverse
Hubble constant, $M_P=\ell_P^{-1}=10^{19}$GeV$=10^{33}$cm is the
Planck mass (or inverse Planck scale, hereafter $c=\hbar=1$).
$\gamma$-function relates the Hubble size of the Universe with its
expansion factor.

\begin{table}
  \caption{Basic cosmological parameters.}
  \label{tab:table1}
  \begin{tabular}{ll}
    \hline
    Hubble parameter                    & $h=0.7$              \\
    CMB temperature                     & $T=2.725 K$          \\
    3-space curvature                   & $\Omega_\kappa=0$    \\
    cosmological density of baryons     & $\Omega_{B}=0.05$    \\
    cosmological density of dark matter & $\Omega_{DM}=0.23 $  \\
    cosmological density of dark energy & $\Omega_\Lambda=0.7$ \\
    power-spectrum index                & $n_S=0.96$           \\
    \hline
  \end{tabular}
\end{table}

Eqs.(3) evidence that all transitions from radiation to matter and
to DE dominated expansions occurred at small energies pretty well
known to atomic physics ($T_{rad}= 2.5\cdot 10^{-4}/a$ eV). When
extrapolating eqs.(3) to earlier times (or higher energies) we
learn the following properties of our Universe:
\begin{itemize}
\item The Universe is large, $(H_0\ell_P)^{-1}\sim 10^{61}$. At
the beginning of the expansion (2) the physical size of the
Universe was a factor $10^{30}$ higher than Planckian size
($a/H_0>\sim$ the current length of relic quanta). Such a big
factor can be explained by a pre-existed short inflationary stage
with $\gamma<1$ (BBS). \item The cosmological perturbations are
uncausal (scales enter horizon at $\gamma >1$). Eqs.(2) describe
decay of $\gamma$ from $2$ to $0.4$. To explain uncausality, one
has to admit a pre-existed period of cosmological expansion with
$\gamma$ rising from values smaller than unity (BBS).
\end{itemize}

Existence of BBS in the early Universe prompts us a solution of DE
problem. Indeed, within $14$ billion years the Universe was at
least twice in state of inflation ($\gamma<1$, by the definition).
There could be more than two stages with $\gamma<1$. Therefore, we
guess that similar physical reasons could be responsible for
different inflationary stages.

Actually, we study the physics of the final period of BBS when
analyzing the large scale structure in the Universe and the
products of BBS inflaton decay (photons, baryons, etc.). On the
other hand, we witness the beginning of the DE {\it stage} of new
inflation. Assuming similar physical reasons causing both stages
(BBS and DES) we come to conclusion that each of these stages must
have the beginning and the end.

Let us illustrate it on the example of simplest inflatons -- {\it
weakly interacting massive field}s \footnote{Recall two important
periods of WIMF evolution: friction domination (the slow-roll
period) and free oscillations (the WIMP period).}. Each
inflationary stage starts with the beginning of appropriate WIMF
domination at slow-roll period of its evolution, and ends by
transferring either to the WIMF oscillations (domination of the
non-relativistic WIMPs) or to the WIMF decay into massless degrees
of freedom.

In this scenario the history of the Universe represents the
history of relaxation of massive fields. How to verify this simple
idea? One of the possibilities can be measuring the $w$ and
$\gamma$ parameters. Say, if DE is a $\phi$-scalar WIMF then the
prediction is $1+w=0.053\cdot(M_P/\phi)^2$, which can be measured
both in sign and value.

\section{In search for DM particles}

DM are WIMPs which were non-relativistic long before the structure
formation in the Universe (back to $T_{rad} > 10$ keV). We do not
know weather WIMPs have decoupled from the thermal bath of
particles or originated evolutionary from some WIMF having never
been in equilibrium with other particles.

Currently, there are several hypotheses on the origin of DM, but
none of them has been confirmed so far. There exist observational
arguments indicating that the DM mystery is related with baryon
asymmetry in the Universe. Two of these arguments are the most
appealing: \begin{itemize} \item {The energy densities of both
non-relativistic components, baryons and DM, are close to each
other now and at the moment of their generation} \item {The
characteristic scales of spatial distributions of baryon and DM
are identical (the cosmological horizon of equal densities of
radiation and matter = the sound horizon of hydrogen
recombination)}
\end{itemize}
However, now there are no generally-accepted theories of the DM
and baryon asymmetry.

Where is dark matter?

We know that luminous constituent of matter is observed as stars
residing in galaxies of different masses and in the form of X-ray
gas in clusters of galaxies. However, a greater amount of ordinary
matter is contained in rarefied intergalactic gas with
temperatures from several to hundred eV and also in MACHO-objects
which are the compact remnants of star evolution and the objects
of small masses. Since these structures mostly have low luminosity
they are traditionally called "dark baryons".

Several scientific groups (MACHO, EROS and others) carried out the
investigation of the number and distribution of compact dark
objects in the halo of our Galaxy, which was based on
micro-lensing events. The combined analysis resulted in an
important bound -- no more than $20\%$ of the entire halo mass is
contained in the MACHO-objects of masses ranging from the Moon to
star masses. The rest of the halo dark matter consists of unknown
particles.

Where else is non-baryonic DM hidden?

The development of high technologies in observational astronomy of
the 20th century allowed us to get a clear-cut answer to this
question -- non-baryonic DM is contained in gravitationally bound
systems (DM halos). DM particles are non-relativistic and weakly
interacting. Unlike baryons, they do not dissipate whereas baryons
are radiationally cooled and settle near the halo centers
attaining rotational equilibrium. DM stays distributed around the
visible matter of galaxies with characteristic scale $\sim200$
kpc. For example, in the Local Group which comprises Andromeda and
Milky Way, more than a half of all DM belongs to these two large
galaxies.

Particles with required properties are absent in the standard
model of particle physics. An important parameter that cannot be
determined from observations due to the Equivalence Principle is
the mass of particle. In this situation the experiment "goes
there, do not know where" (after the Russian fairy tale). The main
candidates are listed in Table \ref{tab:table2} in ascending order
of their rest masses.
\begin{table}
  \caption{Candidates for non-baryonic dark matter particles.}
  \label{tab:table2}
  \begin{tabular}{ll}
    \hline
    candidate                 & mass             \\
    \hline
    gravitons                 & $10^{-21}$eV     \\
      axions                  & $10^{-5}$eV      \\
      "sterile" neutrino      & 10 keV           \\
      mirror matter           & 1 GeV            \\
      neutralino       & 100 GeV          \\
      super-massive particles & $10^{13}$GeV     \\
      monopoles and defects   & $10^{19}$GeV     \\
      primordial black holes  & $10^{-16}-10^{-7}M_\odot$ \\
    \hline
  \end{tabular}
\end{table}

One of the versions on agenda -- the neutralino hypothesis --
rises from minimal supersymmetry. This hypothesis can be verified
in CERN at LHC that will run in 2008. The expected mass of these
particles is $\sim 100~GeV$, and their density in our Galaxy is a
particle per cup of coffee.

DM particles are being searched in many experiments all over the
world. Interestingly, the neutralino hypothesis can be
independently verified both in underground experiments on elastic
scattering and by indirect data on neutralino annihilation in
Galaxy. So far the positive signal has been found only in one of
the underground detectors (DAMA), where a season signal of unknown
origin has been observed for several years now. But the range of
masses and cross-sections associated with this experiment has not
been confirmed in other experiments, which makes reliability and
meaning of the results quite questionable.

Neutralino give an important possibility of indirect detection by
their annihilation gamma-ray flux. During the process of
hierarchic clustering these particles could form mini-halos of the
Earth masses and characteristic sizes comparable to that of the
Solar System. Some of these mini-halos could stay intact till now.
With high probability the Earth itself is inside one of these
halos where the particle density is as much as tens of times
higher than the mean halo density. Hence, the probability of both
direct and indirect detection of DM in our Galaxy gets higher.
Availability of so different search techniques gives a solid hope
that the physical nature of at one version of DM will soon be
verified.

\section{In the beginning was sound}

Let us consider the first order geometry in more detail.

The effect of the quantum-gravitational generation of massless
fields is well-studied. Matter particles can be created with this
effect (see~\cite{ref:GMM, ref:Zeldovich} etc.) (although the
background radiation photons emerged as a result of the BBS
proto-matter decay in the early Universe). The gravitational
waves~\cite{ref:Grischuk} and the density
perturbations~\cite{ref:Lukash-1} are generated in the same way
since they are massless fields and their creation is not
suppressed by the threshold energy condition. The problem of the
vortical perturbation creation is waiting for its researchers.

The theory of the $S$ and $T$ perturbation modes in the Friedmann
Universe reduces to a quantum-mechanical problem of independent
oscillators $q_{k}(\eta)$ in the external parametrical field
$\alpha(\eta)$ in Minkovski space with time coordinate
$\eta=\int{dt/a}$. The action and the Lagrangian of the elementary
oscillators depend on their spatial frequency $k\in(0,\infty)$:
\begin{eqnletter}
\label{eq-2}
S_{k} & = &\displaystyle\int{L_{k}d\eta}, \\
L_{k} & = & \displaystyle\frac{\alpha^{2}}{2k^{3}}\left(
q'^{2}-\omega^{2}q^{2}\right).
\end{eqnletter}
A prime denotes derivative with respect to time $\eta$,
$\omega=\beta k$ is the oscillator frequency, $\beta$ is the speed
of the perturbation propagation in the vacuum-speed-of-light units
(henceforth, the sub-index $k$ for $q$ is omitted). In the case of
the $T$ mode $q\equiv q_{T}$ is a transversal and traceless
component of the metric tensor,
\begin{equation}
\label{eq-3} \alpha^{2}_{T}=\displaystyle\frac{a^{2}}{8\pi G},
\qquad \beta=1.
\end{equation}
In the case of the $S$ mode $q\equiv q_{S}$ is a linear
superposition of the longitudinal gravitational potential (the
scale factor perturbation) and the potential of the 3-velocity of
medium times the Hubble parameter~\cite{ref:Lukash-1}:
\begin{equation}
\label{eq-4} q_{S}=A+H\upsilon, \qquad
\alpha^{2}_{S}=\displaystyle\frac{a^{2}\gamma}{4\pi G\beta^{2}},
\end{equation}
where $A\equiv\delta a/a$, $\upsilon\equiv\delta\phi/{\dot\phi}$.

As it is seen from eq.~(\ref{eq-3}), the field $q_{T}$ is
fundamental, because it is minimally coupled with background
metrics and does not depend on matter properties (in General
Relativity the speed of gravitational waves is equal to the speed
of light). On the contrary, the relation between $q_{S}$ and the
external field~(\ref{eq-4}) is more complicated: it includes both
derivatives of the scale factor and some matter characteristics
(e.g. the speed of perturbation propagation in the medium). We do
not know anything about the proto-matter in the Early Universe.
There are only general suggestions concerning this problem.

Commonly, ideal medium is considered with the energy-momentum
tensor depending on the energy density $\epsilon$, the pressure
$p$, and the 4-velocity $u^{\mu}$. For the $S$ mode, the
4-velocity is potential and represented as a gradient of the
4-scalar $\phi$:
\begin{equation}
\label{eq-5} T_{\mu\nu}=(\epsilon+p)u_{\mu}u_{\nu}-pg_{\mu\nu},
\qquad
u_{\mu}=\phi_{,\mu}/w,
\end{equation}
where a comma denotes the coordinate derivative, and
$w^{2}=\phi_{,\nu}\phi_{,\mu}g^{\mu\nu}$ is a normalizing
function. The speed of sound is given by "equation of state" and
relates comoving perturbations of the pressure and energy density:
\begin{equation}
\label{eq-6} \delta p_{c}=\beta^{2}\delta\epsilon_{c},
\end{equation}
where $\delta X_{c}=\delta X-\upsilon\dot{X}$, and
$\upsilon\equiv\delta\phi/w$ is the potential of the 3-velocity of
medium.

In the linear order of the perturbation theory the ideal medium
concept is equivalent to the field concept where the Lagrangian
density $L=L(w,\phi)$ is ascribed to the material field
$\phi$~\cite{ref:Lukash-1}-~\cite{ref:Strokov}. In the field
approach the speed of the perturbation propagation is found from
equation:
\begin{equation}
\label{eq-7}
\beta^{-2}=\frac{\partial \ln|\partial L/\partial w
|}{\partial\ln|w|},
\end{equation}
which also corresponds to eq.~(\ref{eq-6}). To zeroth order,
$\beta$ is a function time. However, in most models of the early
Universe one usually assumes $\beta\sim 1$ (e.g. at the
radiation-dominated stage $\beta=1/\sqrt{3}$).

The evolution of the elementary oscillators is given by
Klein-Gordon equation:
\begin{equation}
\label{eq-8} \bar{q}''+(\omega^{2}-U)\bar{q}=0,
\end{equation}
where
\begin{equation}
\label{eq-9} \bar{q}\equiv\alpha q, \qquad
U\displaystyle\equiv\frac{\alpha''}{\alpha}.
\end{equation}
The solution of eq.(10) has two asymptotics: an adiabatic one
($\omega^{2}>U$) when the oscillator freely oscillates with the
excitation amplitude being adiabatically damped
($|q|\sim(\alpha\sqrt{\beta})^{-1}$), and a parametric one
($\omega^{2}<U$) when the $q$ field freezes out ($q\to const$).
The latter conditions in respect to quantum field theory implies a
parametrical generation of a pair of particles from the state with
an elementary excitation (see Fig. 2).
\begin{figure}
\includegraphics{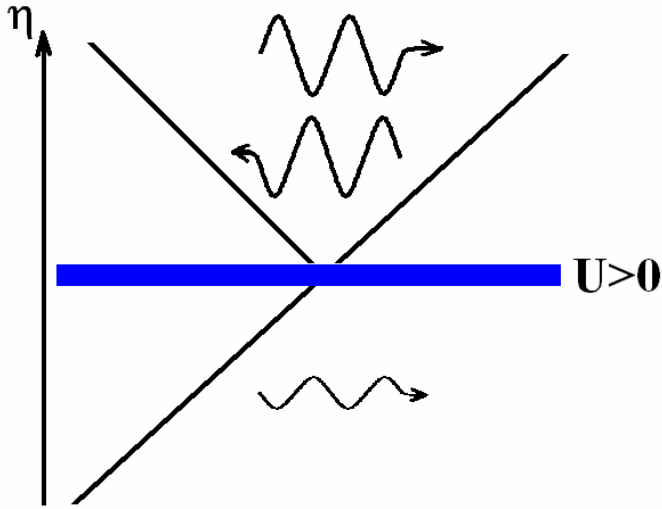}     
\caption{Illustration of solution of scattering problem for eq.
(\ref{eq-8}).}
\end{figure}

Quantitatively, the spectra of the generated perturbations depend
on the initial state of the oscillators:
\begin{equation}
T\equiv 2\langle q_{T}^{2}\rangle, \qquad
S\equiv \langle
q_{S}^{2}\rangle,
\end{equation}
where the field operators are given in the parametrical zone
($q\sim const$). The factor $2$ in the tensor mode expression is
due to two polarizations of gravitational waves. The state
$\langle\rangle$ is considered to be a ground state, i.e. it
corresponds to the minimal level of the initial oscillator
excitation. This is the basic hypothesis of the Big Bang theory.
In case the adiabatic zone is there, the ground (vacuum) state of
the elementary oscillators is unique~\cite{ref:Lukash-3}.

Thus, assuming that the function $U$ grows from zero with time
(i.e. the initial adiabatic zone is followed by the parametric
one) and $\beta\sim 1$, we obtain a universal and general result
for the $T(k)$ and $S(k)$ spectra:
\begin{equation}
\label{eq-11} T\displaystyle=\frac{4\pi
(2-\gamma)H^{2}}{M_{P}^{2}}, \qquad \displaystyle\frac{T}{S}=
4\gamma,
\end{equation}
where $k\simeq aH$ specifies the moment of creation
($\omega^{2}=U$). As it is seen from eq.~(\ref{eq-11}), the theory
does not discriminate the $T$ from $S$ mode. It is the value of
the factor $\gamma$ in the creation period that matters when we
relate $T$ and $S$.

From the observed fact that the $T$ mode is small in our Universe
(see eq. (\ref{eq-1})) we obtain the upper bound on the energetic
scale of the Big Bang and on the parameter $\gamma$ in the early
Universe:
\begin{equation}
\label{eq-12} H<10^{13}GeV, \qquad \gamma<0.05.
\end{equation}
The latter condition implies that BBS was just inflation
($\gamma<1$).

We have important information on phases: the fields are generated
in certain phase, only the growing evolution branch is
parametrically amplified. Let us illustrate it for a scattering
problem, with $U=0$ at the \textit{initial} (adiabatic) and
\textit{final} (radiation-dominated, $a\propto\eta$) evolution
stages (see Fig. 2).

For either of the two above-mentioned stages the general solution
is
\begin{equation}
\bar{q}=C_{1}\sin\omega\eta+C_{2}\cos\omega\eta,
\end{equation}
where the constant operators $C_{1,2}$ yield the amplitudes of the
"growing" and "decaying" solutions. In the vacuum state the
initial time phase is arbitrary:
$\langle|C_{1}^{(in)}|\rangle=\langle|C_{2}^{(in)}|\rangle$.
However, the solution of the evolution equations yields that only
the growing branch of the sound perturbations takes advantage at
the radiation-dominated stage\footnote{This important result can
be explained by the fact that only growing solution is consistent
with the isotropic Friedmannian expansion at small times
($\omega\eta<<1$).}:
$\langle|C_{1}^{(fin)}|\rangle>>\langle|C_{2}^{(fin)}|\rangle$. By
the moment of matter-radiation decoupling at the recombination
era, the radiation spectrum appears modulated with typical phase
scales $k_{n}=n\pi\sqrt{3}/\eta_{rec}$, where $n$ is a natural
number.

It is these acoustic oscillations that are observed in the spectra
of the CMB anisotropy (see Fig. 3, the highest peak corresponds to
$n=1$) and the density perturbations, which confirms the
quantum-gravitational origin of the $S$ mode. We see, the standard
cosmological model can begin as follows. "In the beginning was
sound. And the sound was of the Big Bang". It differs a bit from
the scenario described in the Bible.

\begin{figure}
\includegraphics{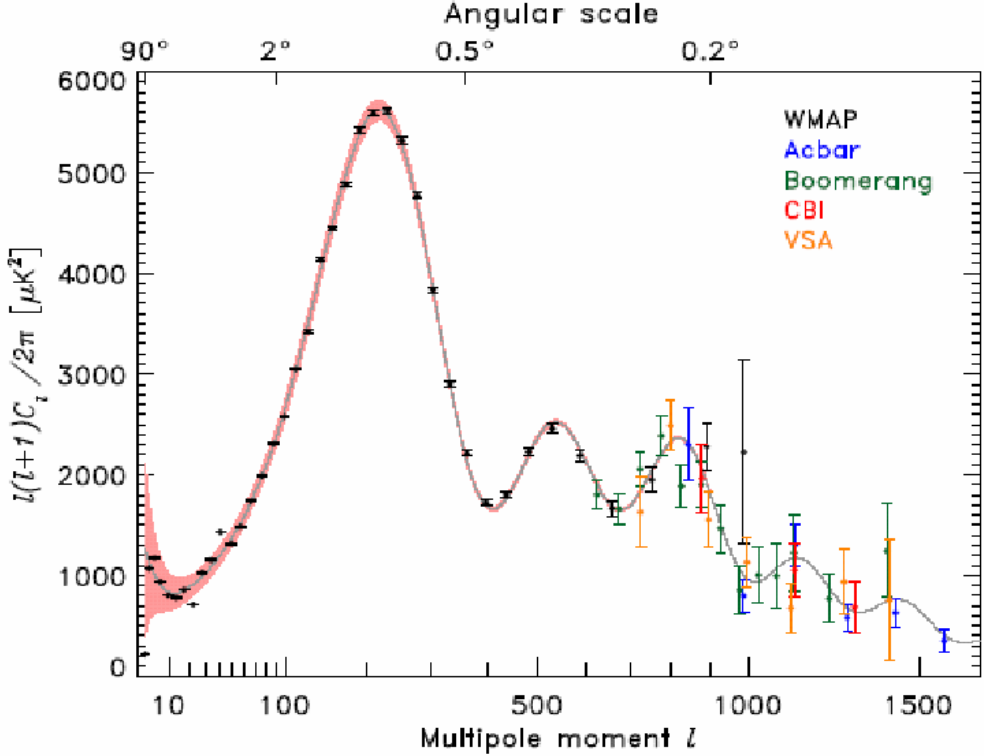}     
\caption{Manifestation of sound modulation in the CMB anisotropy
spectrum.}
\end{figure}

The sound modulation in the density perturbation spectrum is
suppressed by the small factor of the baryon fraction in the
entire budget of matter density. This allows one to determine this
fraction independently of other cosmological tests. The
oscillation scale itself is an example of the standard ruler that
is used to determine cosmological parameters of the Universe. The
problem of degeneracy of cosmological parameters reconstructed
from observational data hindered scientists from building-up the
model of the Universe for many years. But now the acuteness of
this problem is looser thanks to many independent and
complementary observational tests.

To summarize we can say that in principle the problem of the
generation of the primordial cosmological perturbations and the
large scale structure of the Universe is solved today. The theory
of the quantum-gravitational creation of perturbations in the
early Universe will be finally confirmed as soon as the T mode is
discovered, which is anticipated in the nearest future. For
example, the simplest BBS (power-law inflation on massive field)
predicts the T mode amplitude to be only as much as 5 times
smaller than that of the S mode (which corresponds to $\gamma\sim
10^{-2}$)~\cite{ref:LM}. Modern devices and technologies are quite
able to solve the problem of registering such small signals
analyzing observational data on the CMB anisotropy and
polarization.

\section{On the verge of new physics}

Nowadays it became possible to separately determine properties of
the early and late Universe from observational astronomical data.
We understand how the primordial cosmological density
perturbations that formed the structure of Universe emerged. We
know crucial cosmological parameters on which the standard model
of the Universe is based, and the latter has no viable rivals.
However, some fundamental questions of the origin of the Big Bang
and of main matter constituents remain unsolved.

Observational discovery of the tensor mode of the cosmological
perturbations is a key to building-up the model of the early
Universe. In this domain of our knowledge we have a clear-cut
theory prediction that is already verified in the case of the $S$
mode and can be experimentally verified for the $T$ mode in the
nearest future.

Giving a long list of hypothetical possibilities where and how to
look for DM particles and DM physics theory has exhausted itself.
Now it is experiment's turn. The current situation reminds great
moments in history of science when quarks, $W$- and $Z$-bosons,
neutrino oscillations, the CMB anisotropy and polarization were
discovered.

One question is beyond the scope of this review -- why Nature is
so generous and allows us to reveal its secrets?

\acknowledgments This work was partially supported by the Russian
Foundation for Basic Research (07-02-00886).

\end{document}